\documentclass[aip,jcp,amsmath,amssymb,reprint]{revtex4-1}
\usepackage{graphicx}

\usepackage{bm,booktabs}
\usepackage{xcolor}
\usepackage[normalem]{ulem}
\allowdisplaybreaks

\makeatletter
\AtBeginDocument{\immediate\write\@auxout{\string\citation{aipnum41Control}}}
\makeatother

\newcommand{\ad}[1]{\hat a^{\dagger}_{#1}}
\newcommand{\an}[1]{\hat a_{#1}}
\newcommand{\asym}[1]{\underline{#1}}   
\newcommand{\eop}[2]{\hat a^{#1}_{#2}}       

\begin{document}

\title{Construction of downfolded Hamiltonians from projective transcorrelation}

\author{Seiichiro Ten-no}
\email{tenno@garnet.kobe-u.ac.jp}
\affiliation{Graduate School of System Informatics, Kobe University,
             Nada-ku, Kobe 657-8501, Japan}

\begin{abstract}
Projective transcorrelation recovers the short-range electron correlation by a similarity transformation with a geminal function $f(r_{12})$, at the cost of an effective Hamiltonian containing a 3-body term.
We replace the term by an effective operator of rank at most two, obtained from the two-body cumulant (2C) approximation to the three-particle reduced density matrix.
The 2C 3-body energy is written without cumulants, and the effective one- and two-body interactions are derived as its partial derivatives with respect to the reduced density matrices.
The truncation is assessed on the bond dissociation of N$_2$, the atomization and reaction energies of the HEAT set with CCSD(T), and the CAS-pTC model, whose downfolded Hamiltonian is held on a qubit register with the number of Pauli strings reduced from the sixth power of the orbital count to the fourth.
\end{abstract}

\maketitle

\section{Introduction}

The Coulomb singularity of the electron--electron interaction imprints the cusp behavior on the exact wavefunction\cite{kato1957,pack1966} that orbital expansions reproduce very slowly, and the resulting $\ell^{-3}$ convergence of the correlation energy with the basis-set angular momentum\cite{kutzelnigg1992} is the practical limit on the accuracy of wavefunction theory.
Explicitly correlated F12 theory augments the wavefunction with terms containing $f(r_{12})$ and has become a widely adopted standard tool, offering an accurate and robust treatment of complex systems.\cite{klopper2006,helgaker2008,tenno2012wires,tenno2012tca,hattig2012,kong2012,shiozaki2013,grueneis2017,ma2018} 

Another approach is transcorrelation that removes the singularity from the Hamiltonian by the similarity transformation $\bar H=e^{-\hat J}\hat H\,e^{\hat J}$ with a Jastrow correlator.\cite{hirschfelder1963,boysHandy1969a,boysHandy1969}
Since $e^{\hat J}$ is nonunitary, the non-Hermitian transformed Hamiltonian loses the variational bound.
The early work concentrated on minimizing the variance instead.\cite{handy1971}
Interest has revived recently with frozen Gaussian geminals,\cite{tenno2000a} biorthogonal formulations,\cite{hino2001,hino2002} applications to the electron gas,\cite{armour1980,umezawa2004,luo2012} solids,\cite{sakuma2006,ochi2012,ochi2014a,ochi2014b,luo2026} and, with Jastrow factors optimized for the purpose,\cite{dobrautz2022,haupt2023,haupt2026} in combination with essentially every modern solver for strong correlation, full configuration interaction (FCI) quantum Monte Carlo,\cite{cohen2019} selected configuration interaction,\cite{ammar2022,ammar2022b,ammar2023,ammar2024} the density matrix renormalization group (DMRG),\cite{baiardi2020,baiardi2022,liao2023} and quantum computing.\cite{mcardle2020,motta2020,schleich2022,dobrautz2024,magnusson2024}
Some of these applications employ the canonical transcorrelation which is a unitary variant built on the F12 ansatz rather than a Jastrow factor.\cite{yanai2012,masteran2025}
More recently, we have developed the projective transcorrelation (pTC) that features a terminating series and the satisfaction of $s$- and $p$-wave cusp conditions.\cite{tenno2023ptc}
The former follows from a generator containing creation operators in the complement of the orbital basis, which terminates the expansion at the double commutator.
The latter is from the SP ansatz with a Slater-type correlation factor.\cite{tenno2004jcp,tenno2004cpl}

All of these methods share a common problem, {\it i.e.} the treatment of many-electron interactions.
A similarity transformation generated by a two-body operator produces, through the double commutator, an effective Hamiltonian including 3-body interactions and more in general.
The evaluation of many-electron integrals is well established in F12 theory by using the resolution of the identity (RI),\cite{kutzelnigg1991} auxiliary basis set (ABS),\cite{klopper2002} complementary ABS (CABS),\cite{valeev2004} or numerical quadratures.\cite{tenno2004jcp,tenno2007}
Even when the integrals are inexpensive, the 3-body operator has $M^{6}$ elements over $M$ orbitals, which exceeds available memory for any production basis, and ordinary many-body solvers such as coupled cluster (CC) cannot accommodate many-electron integrals explicitly.

The frequently used tool is the cumulant expansion of the reduced density matrices.
The reconstruction of the three- and four-particle RDMs from the lower ones was introduced in the work on the contracted Schr\"odinger equation.\cite{colmenero1993a,colmenero1993,nakatsuji1996,mazziotti1998}
Kutzelnigg and Mukherjee gave the cumulant expansion of the reduced density matrices, defining the cumulants as the connected parts of the RDMs and constructing the generalized normal ordering whose contractions are the cumulants.\cite{KM1997,KM1999}
The same reconstruction has made canonical transformation\cite{yanaiChan2006,neuscamman2010jcp,neuscamman2010rev} and multireference perturbation theory\cite{kurashige2011,sivalingam2016} feasible with DMRG references.
For transcorrelation, the 3-body term has been contracted to two-body form in the xTC method, which normal-orders it with respect to a single determinant,\cite{christlmaier2023} and the resulting Hamiltonian has been combined with coupled cluster\cite{schraivogel2021,schraivogel2023} and corrected additively at the reference level.\cite{hauskrecht2026}

Explicitly correlated approaches treat the short-range correlation directly, which keeps the orbital space small.
The terminology of downfolding is used in the electronic structure theory of solids for the construction of a low-energy effective model that absorbs the high-energy degrees of freedom.\cite{aryasetiawan2004,imada2010}
The present work shares the same purpose, with the high angular momentum part of the orbital basis playing the role of the high-energy bands.
Because the geminal holds the cusp explicitly, pTC returns an effective Hamiltonian already close to the complete basis set (CBS) limit in a modest orbital basis.
FCI and its approximations --- selected configuration interaction,\cite{huron1973,holmes2016} selected coupled cluster,\cite{xu2018,xu2020} quantum Monte Carlo in configuration space\cite{booth2009,tenno2013} and DMRG\cite{white1992,white1999} --- still scale steeply with the size of the orbital space.
The reduction of the orbital space at a given accuracy by pTC therefore lowers the cost of a correlation treatment.
This gain is even sharper in quantum computing, where each spin orbital costs a qubit, and transcorrelation has been shown to reach a given accuracy at a fixed register size.\cite{motta2020,kumar2022,sokolov2023,dobrautz2024}
On a near-term register, the energy is accumulated from sampled expectation values, so that the orbital count is paid in qubits and the many-body rank in the number of terms to be measured.

In the present work, we discuss the construction of downfolded Hamiltonians of pTC in view of the cumulant expansion of the 3-body interactions and of necessary corrections. 
The two-body cumulant (2C) approximation discards only the connected three-particle correlation and effective one- and two-body interactions are obtained as partial derivatives of the 3-body energy with respect to the one- and two-particle densities.
The approximation is assessed on the bond dissociation of N$_2$, the HEAT set with CCSD(T)-pTC, and a qubit register under the Jordan--Wigner transformation.

\section{Theory}

\subsection{The projective transcorrelated Hamiltonian}
\label{sec:ptc}

Transcorrelation replaces the Schr\"odinger equation by its similarity transform with a generator containing a correlation factor.
In pTC,\cite{tenno2023ptc} the transformed Hamiltonian is
\begin{align}
\hat{\mathcal H}_{\rm pTC}
   =\mathcal B\{e^{-\hat{\mathcal G}}\,\hat H\,e^{\hat{\mathcal G}}\},
\label{eq:simtrans}
\end{align}
where $\mathcal B\{\cdots\}$ takes an operator defined in the CBS and returns its second-quantized representation within the given basis set (GBS).
The generator is
\begin{align}
\hat{\mathcal G}=\frac12\sum_{\kappa\lambda rs}
  \langle\kappa\lambda|\hat Q_{12}\hat{\mathcal R}_{12}|rs\rangle\,
  \eop{\kappa\lambda}{rs},
\label{eq:gen}
\end{align}
with the replacement operators
\begin{align}
\eop{pq\cdots}{rs\cdots}=\ad{p}\ad{q}\cdots\an{s}\an{r}
\label{eq:excop}
\end{align}
over spin orbitals.
Orbital indices follow Ref.~\onlinecite{tenno2023ptc}: $p,q,\dots$ label the GBS (projector $\hat P_{n}$), $\alpha,\beta,\dots$ its orthogonal complement ($\hat Q_{n}$), $\kappa,\lambda,\dots$ the complete basis set ($1=\hat P_{n}+\hat Q_{n}$), and $i,j,\dots$ occupied and $a,b,\dots$ virtual orbitals.
The spatial part of a spin-orbital is shown separately by an upper-case label with the spin explicitly, $p=P\sigma$ with $\sigma=\alpha,\beta$.
The projector
\begin{align}
\hat Q_{12}=(1-\hat P_{1}\hat P_{2})\hat {\mathcal S}_{12},
\label{eq:q12}
\end{align}
removes the geminal contributions already representable in the GBS and $\hat {\mathcal S}_{12}$ is a supplementary operator to uncorrelate frozen core orbitals.\cite{tenno2023ptc}
For the rational generator $\hat{\mathcal R}_{12}$, we use the fixed-amplitude (SP) ansatz\cite{tenno2004jcp} with the Slater-type correlation factor\cite{tenno2004cpl}
\begin{align}
\hat{\mathcal R}_{12}&=f(r_{12})(\tfrac38+\tfrac18\hat P_{12}),\\
f(r_{12})&=-\gamma^{-1}\exp(-\gamma r_{12}),
\label{eq:sp}
\end{align}
in order to satisfy the $s$- and $p$-wave cusp conditions simultaneously, where $\hat P_{12}$ interchanges the spatial coordinates and leaves the spins untouched like
\begin{align}
\hat P_{12}|P\sigma_1 Q\sigma_2\rangle=|Q\sigma_1 P\sigma_2 \rangle.
\label{eq:p12}
\end{align}
Unlike a Jastrow factor, $\hat{\mathcal G}$ contains at least one creation operator in the complementary space, which can contribute only when contracted with $\hat H$ through $\hat Q_{n}$.
The expansion therefore terminates at the double commutator,
\begin{align}
\hat{\mathcal H}_{\rm pTC}
  &=\mathcal B\{\hat H+[\hat H,\hat{\mathcal G}]
     +\tfrac12[[\hat H,\hat{\mathcal G}],\hat{\mathcal G}]\}\nonumber\\
  &=\hat{\mathcal H}+\hat{\mathcal H}_{\rm 2h}+\hat{\mathcal H}_{\rm 2w}
   +\hat{\mathcal H}_{\rm 3l}+\hat{\mathcal H}_{\rm 3q}+\hat{\mathcal H}_{\rm 4},
\label{eq:hptc}
\end{align}
where $\hat{\mathcal H}=\mathcal B\{\hat H\}$, $\hat{\mathcal H}_{\rm 2h}$ and $\hat{\mathcal H}_{\rm 2w}$ are the two-body terms of the single and double contractions of the Hamiltonian and $\hat {\mathcal G}$, respectively, and $\hat{\mathcal H}_{\rm 3l}$ and $\hat{\mathcal H}_{\rm 3q}$ the 3-body terms linear and quadratic in $\hat{\mathcal G}$, and $\hat{\mathcal H}_{\rm 4}$ the four-body term.
Explicitly, the three terms linear in $\hat{\mathcal G}$ are
\begin{align}
\hat{\mathcal H}_{\rm 2h}&=\sum_{pqrs}
  \langle pq|\hat h_{1}(1-\hat P_{1})\hat{\mathcal S}_{12}\hat{\mathcal R}_{12}|rs\rangle
  \,\eop{pq}{rs},
\label{eq:h2h}\\[2pt]
\hat{\mathcal H}_{\rm 2w}&=\frac12\sum_{pqrs}
  \langle pq|r_{12}^{-1}\hat Q_{12}\hat{\mathcal R}_{12}|rs\rangle
  \,\eop{pq}{rs},
\label{eq:h2w}\\[2pt]
\hat{\mathcal H}_{\rm 3l}&=\sum_{pqrstu} L^{pqr}_{stu}\,\eop{pqr}{stu},
\label{eq:h3l}\\[2pt]
L^{pqr}_{stu}&=\langle pqr|r_{12}^{-1}(1-\hat P_{1})
  \hat{\mathcal S}_{13}\hat{\mathcal R}_{13}|stu\rangle .
\label{eq:Ldef}
\end{align}
$\langle\hat{\mathcal H}_{\rm 3q}\rangle$ and $\langle\hat{\mathcal H}_{\rm 4}\rangle$, which are third-order contributions, are usually neglected.
Truncation of Eq.~\eqref{eq:hptc} after the terms linear in $\hat{\mathcal G}$ leaves
\begin{align}
\hat{\mathcal H}_{\rm pTCl}
   =\hat{\mathcal H}+\hat{\mathcal H}_{\rm 2h}+\hat{\mathcal H}_{\rm 2w}
    +\hat{\mathcal H}_{\rm 3l},
\label{eq:hptcl}
\end{align}
which is the Hamiltonian considered in this work.
This form inherits the properties that distinguish pTC from the Jastrow transcorrelation,\cite{tenno2023ptc} {\it i.e.} no spin contamination arises within the non-relativistic framework, the singlet and triplet first-order cusp conditions are satisfied simultaneously, and correlated orbitals may be selected without restriction like the frozen core approximation.
The operator is non-Hermitian, and higher-order explicitly correlated contributions are incorporated automatically once it enters a correlated calculation.

\subsection{Effective mean field for $\hat{\mathcal H}_{\rm 3l}$ using the
         cumulant expansion}
\label{sec:cumulant}
\label{sec:meanfield}

The 3-electron integrals are usually too numerous to be accommodated explicitly, and we explore a low-rank approximation using the cumulant expansion.
The energy due to Eq.~\eqref{eq:h3l} is
\begin{align}
\langle\hat{\mathcal H}_{\rm 3l}\rangle
  =\sum_{pqrstu} L^{pqr}_{stu}\,\gamma^{pqr}_{stu},
\label{eq:E3l}
\end{align}
where the reduced density matrices (RDMs) of the reference state $|\Psi\rangle$ are, with the replacement operators of Eq.~\eqref{eq:excop},
\begin{align}
\gamma^{pq\cdots}_{rs\cdots}
  =\langle\Psi|\eop{pq\cdots}{rs\cdots}|\Psi\rangle
  =\langle\ad{p}\ad{q}\cdots\an{s}\an{r}\rangle.
\label{eq:gn}
\end{align}
Each RDM is antisymmetric under permutation of its upper labels and, separately, of its lower labels.

The cumulants $\lambda^{(n)}$ are defined\cite{KM1997,KM1999} by requiring $\gamma^{(n)}$ to be the sum, over all partitions of the $n$ index pairs into disjoint blocks, of products of the cumulants of the blocks, each term signed by the parity of the permutation that assigns lower to upper labels.
Thus $\lambda^{(n)}$ is the fully connected part of $\gamma^{(n)}$.
The first three relations read
\begin{align}
\gamma^{p}_{s}&=\lambda^{p}_{s},\label{eq:so1}\\
\gamma^{pq}_{st}&=\asym{\gamma^{p}_{s}\gamma^{q}_{t}}+\lambda^{pq}_{st},
\label{eq:so2}\\
\gamma^{pqr}_{stu}&=\asym{\gamma^{p}_{s}\gamma^{q}_{t}\gamma^{r}_{u}}
  +\asym{\gamma^{p}_{s}\lambda^{qr}_{tu}}+\lambda^{pqr}_{stu},
\label{eq:so3}
\end{align}
where $\asym{X}$ stands for the sum of all signed antisymmetric combinations of the labels of $X$ as
\begin{align}
\asym{\gamma^{p}_{s}\gamma^{q}_{t}}
  &=\gamma^{p}_{s}\gamma^{q}_{t}-\gamma^{p}_{t}\gamma^{q}_{s},
\label{eq:A11}\\
\asym{\gamma^{p}_{s}\lambda^{qr}_{tu}}
  &=\gamma^{p}_{s}\lambda^{qr}_{tu}-\gamma^{p}_{t}\lambda^{qr}_{su}
    +\gamma^{p}_{u}\lambda^{qr}_{st}\nonumber\\
  &\quad-\gamma^{q}_{s}\lambda^{pr}_{tu}+\gamma^{q}_{t}\lambda^{pr}_{su}
    -\gamma^{q}_{u}\lambda^{pr}_{st}\nonumber\\
  &\quad+\gamma^{r}_{s}\lambda^{pq}_{tu}-\gamma^{r}_{t}\lambda^{pq}_{su}
    +\gamma^{r}_{u}\lambda^{pq}_{st},
\label{eq:A21}\\
\asym{\gamma^{p}_{s}\gamma^{q}_{t}\gamma^{r}_{u}}
  &=\gamma^{p}_{s}\gamma^{q}_{t}\gamma^{r}_{u}
    -\gamma^{p}_{s}\gamma^{q}_{u}\gamma^{r}_{t}-\gamma^{p}_{t}\gamma^{q}_{s}\gamma^{r}_{u}
    +\gamma^{p}_{t}\gamma^{q}_{u}\gamma^{r}_{s}\nonumber\\
  &\quad+\gamma^{p}_{u}\gamma^{q}_{s}\gamma^{r}_{t}
    -\gamma^{p}_{u}\gamma^{q}_{t}\gamma^{r}_{s}.
\label{eq:A111}
\end{align}

We now introduce the 2C approximation using the single truncation, $\lambda^{pqr}_{stu}=0$.
As $\gamma^{(3)}$ expressed through $\gamma^{(2)}$ and $\gamma^{(1)}$ without cumulants is needed for implementation, substituting Eq.~\eqref{eq:so2} for $\lambda^{qr}_{tu}$ in Eq.~\eqref{eq:so3} and collecting terms gives
\begin{align}
\gamma^{pqr}_{stu}
\overset{\text{2C}}{\simeq} \asym{\gamma^{p}_{s}\gamma^{qr}_{tu}}
 -2\,\asym{\gamma^{p}_{s}\gamma^{q}_{t}\gamma^{r}_{u}}.
\label{eq:2Crdm-so}
\end{align}
We also define the cruder one-body cumulant (1C) approximation,
\begin{align}
\gamma^{pqr}_{stu}
 \overset{\text{1C}}{\simeq}\asym{\gamma^{p}_{s}\gamma^{q}_{t}\gamma^{r}_{u}}.
\label{eq:1C}
\end{align}
In the Hartree--Fock (HF) limit, $\lambda^{(2)}$ vanishes, and Eqs.~\eqref{eq:2Crdm-so} and \eqref{eq:1C} coincide and both become exact.

Symmetrizing $L^{pqr}_{stu}$ with respect to the 3-electronic coordinates,
\begin{align}
L'^{pqr}_{stu}=\tfrac16(
  &L^{pqr}_{stu}+L^{prq}_{sut}+L^{qpr}_{tsu}\nonumber\\
  &+L^{qrp}_{tus}+L^{rpq}_{ust}+L^{rqp}_{uts}),
\label{eq:Lbar}
\end{align}
the 2C expectation value is expressed as
 \begin{align}
\langle\hat{\mathcal H}_{\rm 3l}\rangle_{\rm 2C}
 =\sum_{pqrstu}\gamma^{p}_{s}
   (\tfrac32\gamma^{qr}_{tu}-2\gamma^{q}_{t}\gamma^{r}_{u})
   \bar L'^{pqr}_{stu},
\label{eq:E2Cbar}
\end{align}
where the overbar denotes antisymmetrization,
\begin{align}
\bar L'^{pqr}_{stu}=
  L'^{pqr}_{stu}-L'^{pqr}_{tsu}-L'^{pqr}_{uts}\nonumber \\
  - L'^{pqr}_{sut}+L'^{pqr}_{tus}+L'^{pqr}_{ust}.
\label{eq:Lbarbar}
\end{align}

The effective one- and two-body operators are generated by the partial derivatives of Eq.~\eqref{eq:E2Cbar} with respect to the RDMs,
\begin{align}
\hat{\mathcal H}^{\rm (eff)}_{1}&=\sum_{ps}
  \frac{\partial\langle\hat{\mathcal H}_{\rm 3l}\rangle_{\rm 2C}}
       {\partial\gamma^{p}_{s}}\;\eop{p}{s},\\
\hat{\mathcal H}^{\rm (eff)}_{2}&=\sum_{pqst}
  \frac{\partial\langle\hat{\mathcal H}_{\rm 3l}\rangle_{\rm 2C}}
       {\partial\gamma^{pq}_{st}}\;\eop{pq}{st},
\label{eq:eff12}
\end{align}
which evaluate to
\begin{align}
\frac{\partial\langle\hat{\mathcal H}_{\rm 3l}\rangle_{\rm 2C}}
     {\partial\gamma^{p}_{s}}
 &=\sum_{qrtu}(\tfrac32\gamma^{qr}_{tu}-6\gamma^{q}_{t}\gamma^{r}_{u})
   \bar L'^{pqr}_{stu},
\label{eq:h1}\\
\frac{\partial\langle\hat{\mathcal H}_{\rm 3l}\rangle_{\rm 2C}}
     {\partial\gamma^{qr}_{tu}}
 &=\tfrac32\sum_{ps}\gamma^{p}_{s}\,\bar L'^{pqr}_{stu}.
\label{eq:h2}
\end{align}
Letting the expectation value be identical to $\langle\hat{\mathcal H}_{\rm 3l}\rangle_{\rm 2C}$, the effective operator beomes
\begin{align}
\hat{\mathcal H}_{\rm 3l}\overset{\text{2C}}{\simeq}
  \hat{\mathcal H}^{\rm (eff)}_{1}+\hat{\mathcal H}^{\rm (eff)}_{2}
  -\langle\hat{\mathcal H}_{\rm 3l}\rangle_{\rm 2C}
  +2\langle\hat{\mathcal H}_{\rm 3l}\rangle_{\rm 1C}
\label{eq:meanfield}
\end{align}
with $\langle\hat{\mathcal H}_{\rm 3l}\rangle_{\rm 1C} =\sum\gamma^{p}_{s}\gamma^{q}_{t}\gamma^{r}_{u}\bar L'^{pqr}_{stu}$.
The traces of the cumulants are useful measures of the accuracy of the effective mean-field,
\begin{align}
\mathrm{Tr}\{\lambda^{(1)}\}&=N,
\label{eq:trlam1}\\
\mathrm{Tr}\{\lambda^{(2)}\}&=\sum_{pq}\gamma^{p}_{q}\gamma^{q}_{p}-N,
\label{eq:trlam2}\\
\mathrm{Tr}\{\lambda^{(3)}\}&=4\sum_{pqr}\gamma^{p}_{q}\gamma^{q}_{r}\gamma^{r}_{p}
 -6\sum_{pq}\gamma^{p}_{q}\gamma^{q}_{p}+2N .
\label{eq:trlam3}
\end{align}
Both $\mathrm{Tr}\{\lambda^{(2)}\}$ and $\mathrm{Tr}\{\lambda^{(3)}\}$ vanish for an idempotent $\gamma^{(1)}$, so that the 1C and 2C approximations coincide at the Hartree--Fock level.

\subsection{Correction of the generalized Brillouin condition}
\label{sec:gbc}

Both $\hat{\mathcal H}_{\rm 2h}$ and $\hat{\mathcal H}_{\rm 3l}$ reach the complementary space through the single projector $1-\hat P_{1}$.
Resolving it as $\sum_{\alpha}|\alpha\rangle\langle\alpha|$, Eqs.~\eqref{eq:h2h} and~\eqref{eq:Ldef} read
\begin{align}
\hat{\mathcal H}_{\rm 2h}&=\sum_{pqrs}\sum_{\alpha}
  \langle p|\hat h|\alpha\rangle\,R^{\alpha q}_{rs}\,\eop{pq}{rs},
\label{eq:h2hcabs}\\
\hat{\mathcal H}_{\rm 3l}&=\sum_{pqrstu}\sum_{\alpha}
  G^{pq}_{\alpha t}\,R^{\alpha r}_{su}\,\eop{pqr}{stu},
\label{eq:Lcabs}
\end{align}
with $G^{pq}_{rs}=\langle pq|r_{12}^{-1}|rs\rangle$ and $R^{pq}_{rs}=\langle pq|\hat{\mathcal S}_{12}\hat{\mathcal R}_{12}|rs\rangle$.
$\hat{\mathcal H}_{\rm 3l}$ separates into the reference mean field and the remainder,\cite{tenno2023ptc}
\begin{align}
\hat{\mathcal H}_{\rm 3l}&=\hat{\mathcal H}_{\rm 2v}
   +[\hat{\mathcal H}_{\rm 3l}]_{r},
\label{eq:split3l}\\
\hat{\mathcal H}_{\rm 2v}&=\sum_{pqrs}\sum_{\alpha}
  \langle p|\hat v^{\rm HF}|\alpha\rangle\,R^{\alpha q}_{rs}\,\eop{pq}{rs}.
\label{eq:h2v}
\end{align}
The mean field arises from $\sum_{k}\bar G^{pk}_{\alpha k}=\langle p|\hat v^{\rm HF}|\alpha\rangle$.
In the single-reference case, the leading term of the perturbation expansion is the expectation value over the HF determinant,
\begin{align}
\langle\hat{\mathcal H}_{\rm 2h}\rangle_{0}
 &=\sum_{ij}\sum_{\alpha}\langle i|\hat h|\alpha\rangle\,\bar R^{\alpha j}_{ij},
\label{eq:e2h}\\
\langle\hat{\mathcal H}_{\rm 2v}\rangle_{0}
 &=\sum_{ij}\sum_{\alpha}\langle i|\hat v^{\rm HF}|\alpha\rangle\,
   \bar R^{\alpha j}_{ij},
\label{eq:2Vperm}\\
\langle[\hat{\mathcal H}_{\rm 3l}]_{r}\rangle_{0}
 &=-\sum_{ijk}\sum_{\alpha}G^{ij}_{\alpha k}\,\bar R^{\alpha k}_{ij}.
\label{eq:3Rperm}
\end{align}
F12 methods usually do not estimate $\langle\hat{\mathcal H}_{\rm 2h}\rangle_{0}$ and $\langle\hat{\mathcal H}_{\rm 2v}\rangle_{0}$ because of the generalized Brillouin condition (GBC),
\begin{align}
\langle i|\hat f|\alpha\rangle\overset{\text{GBC}}{\simeq}0,
\label{eq:gbc}
\end{align}
for $\hat f=\hat h+\hat v^{\rm HF}$.
The GBC is rapidly satisfied as the basis set is enlarged.
The remainder $\langle[\hat{\mathcal H}_{\rm 3l}]_{r}\rangle_{0}$, together with $\langle\hat{\mathcal H}_{\rm 2w}\rangle_{0}$, constitutes the V-term of the F12 correction.
The basis-set convergence of $\langle[\hat{\mathcal H}_{\rm 3l}]_{r}\rangle_{0}$ is much slower than that of the mean-field part $\langle\hat{\mathcal H}_{\rm 2v}\rangle_{0}$.
The number of occupied indices in the objects contracted with $\bar R$ shows that the expansion in $\alpha$ requires angular momentum up to $\ell_{\rm occ}$ for $\langle\hat{\mathcal H}_{\rm 2v}\rangle_{0}$ but up to $3\ell_{\rm occ}$ for $\langle[\hat{\mathcal H}_{\rm 3l}]_{r}\rangle_{0}$ in the one-center (atomic) case, where $\ell_{\rm occ}$ is the highest angular momentum among the occupied orbitals.
Nevertheless, the GBC error embedded in $\hat{\mathcal H}_{\rm pTC}$ becomes significant in a small basis set and cannot easily be removed once the Hamiltonian has been constructed.
In this case, we use the GBC correction
\begin{align}
\Delta_{\rm GBC}=-\langle\hat{\mathcal H}_{\rm 2h}\rangle_{0}-\langle\hat{\mathcal H}_{\rm 2v}\rangle_{0},
\label{eq:dgbc}
\end{align}
and the basis-set incompleteness of the reference function is compensated separately by an HF calculation in a large basis.
The CABS singles correction\cite{adler2007} serves the same purpose, though the cost of an HF calculation is usually negligible compared with that of the subsequent post-HF treatment.

\subsection{Spin-free adaptations}
\label{sec:spinfree}

The effective mean field of the cumulant expansion presented above becomes spin-dependent for open-shell systems, which costs memory and arithmetic even when the spatial orbitals are common to both spins.
We therefore explore the spin-free adaptation of the present pTC framework.
The expectation value of $\hat{\mathcal H}_{\rm 3l}$ is written in the spin-free form as
\begin{align}
\langle\hat{\mathcal H}_{\rm 3l}\rangle
 =\sum_{PQRSTU}L^{PQR}_{STU}\,\Gamma^{PQR}_{STU},
\label{eq:E3sf}
\end{align}
with the spin-free reduced density matrices
\begin{align}
\Gamma^{PQ\cdots}_{RS\cdots}
 &=\langle\Psi|\hat E^{PQ\cdots}_{RS\cdots}|\Psi\rangle
  =\sum_{\sigma\tau\cdots}\gamma^{P\sigma Q\tau\cdots}_{R\sigma S\tau\cdots},
\nonumber\\
\hat E^{PQ\cdots}_{RS\cdots}
 &=\sum_{\sigma\tau\cdots}\eop{P\sigma Q\tau\cdots}{R\sigma S\tau\cdots},
\label{eq:Gdef}
\end{align}
where $L^{PQR}_{STU}$ are integrals over spatial orbitals.

Kutzelnigg, Shamasundar and Mukherjee constructed the cumulant expansion directly for the spin-free RDMs.\cite{KM2010}
With a spin-restricted one-particle density,
\begin{align}
\gamma^{P\sigma}_{Q\tau}=\tfrac12\,\delta_{\sigma\tau}\,\Gamma^{P}_{Q},
\label{eq:g1half}
\end{align}
the spin summation leaves a direct contraction unchanged, $\sum_{\sigma\tau}\gamma^{P\sigma}_{R\sigma}\gamma^{Q\tau}_{S\tau}=\Gamma^{P}_{R}\Gamma^{Q}_{S}$, and halves an exchange, $\sum_{\sigma\tau}\gamma^{P\sigma}_{S\tau}\gamma^{Q\tau}_{R\sigma}=\tfrac12\Gamma^{P}_{S}\Gamma^{Q}_{R}$.
The spin summation refers to the $M_{S}$-averaged ensemble.
Except for a singlet, Eqs.~\eqref{eq:lam2sf}--\eqref{eq:meanfieldsf} are therefore not equivalent to the cumulant expansion of the spin-dependent formulation.
Consequently, the 2- and 3-RDMs are
\begin{align}
\Gamma^{PQ}_{RS}=\Gamma^{P}_{R}\Gamma^{Q}_{S}
 -\tfrac12\,\Gamma^{P}_{S}\Gamma^{Q}_{R}+\Lambda^{PQ}_{RS},
\label{eq:lam2sf}
\end{align}
and
\begin{align}
\Gamma^{PQR}_{STU}
&=\Gamma^{P}_{S}\Lambda^{QR}_{TU}+\Gamma^{Q}_{T}\Lambda^{PR}_{SU}
    +\Gamma^{R}_{U}\Lambda^{PQ}_{ST}\nonumber\\
   &-\tfrac12(\Gamma^{P}_{T}\Lambda^{QR}_{SU}+\Gamma^{Q}_{S}\Lambda^{PR}_{TU}
                 +\Gamma^{Q}_{U}\Lambda^{PR}_{ST}\nonumber\\
   &+\Gamma^{R}_{T}\Lambda^{PQ}_{SU}+\Gamma^{P}_{U}\Lambda^{QR}_{TS}
                 +\Gamma^{R}_{S}\Lambda^{PQ}_{UT})\nonumber\\
   &+\Gamma^{P}_{S}\Gamma^{Q}_{T}\Gamma^{R}_{U}
    -\tfrac12(\Gamma^{P}_{T}\Gamma^{Q}_{S}\Gamma^{R}_{U}\nonumber\\
   &+\Gamma^{P}_{S}\Gamma^{Q}_{U}\Gamma^{R}_{T}
    +\Gamma^{P}_{U}\Gamma^{Q}_{T}\Gamma^{R}_{S})\nonumber\\
   &+\tfrac14(\Gamma^{P}_{T}\Gamma^{Q}_{U}\Gamma^{R}_{S}
    +\Gamma^{P}_{U}\Gamma^{Q}_{S}\Gamma^{R}_{T})
    +\Lambda^{PQR}_{STU}.
\label{eq:lam3sf}
\end{align}
Dropping $\Lambda^{PQR}_{STU}$ and substituting $\Lambda^{PQ}_{RS}$ of Eq.~\eqref{eq:lam2sf} into the above, we obtain the spin-free variant of the 2C approximation,%
\begin{align}
\langle\hat{\mathcal H}_{\rm 3l}\rangle_{\rm 2C}
 =\sum_{PQRSTU}&\Gamma^{P}_{S}[
 (3\Gamma^{QR}_{TU}-2\Gamma^{Q}_{T}\Gamma^{R}_{U})\tilde L'^{PQR}_{STU}\nonumber\\
 &+\Gamma^{Q}_{T}\Gamma^{R}_{U}\,\tilde L'^{PQR}_{SUT}],
\label{eq:E3sf2C}
\end{align}
where
\begin{align}
\tilde L'^{PQR}_{STU}=L'^{PQR}_{STU}-L'^{PQR}_{TSU}.
\label{eq:Ltilde}
\end{align}
As in the spin-dependent case, the effective interactions are obtained by partial differentiation with respect to $\Gamma$,
\begin{align}
\hat{\mathcal H}^{\rm (eff)}_{1}&=\sum_{PS}
  \frac{\partial\langle\hat{\mathcal H}_{\rm 3l}\rangle_{\rm 2C}}
       {\partial\Gamma^{P}_{S}}\;\hat E^{P}_{S},
\label{eq:eff1sf}\\
\hat{\mathcal H}^{\rm (eff)}_{2}&=\sum_{PQST}
  \frac{\partial\langle\hat{\mathcal H}_{\rm 3l}\rangle_{\rm 2C}}
       {\partial\Gamma^{PQ}_{ST}}\;\hat E^{PQ}_{ST},
\label{eq:eff2sf}
\end{align}
with
\begin{align}
\frac{\partial\langle\hat{\mathcal H}_{\rm 3l}\rangle_{\rm 2C}}
     {\partial\Gamma^{P}_{S}}
 &=\sum_{QRTU}[(3\Gamma^{QR}_{TU}-6\Gamma^{Q}_{T}\Gamma^{R}_{U}) \tilde L'^{PQR}_{STU}\nonumber\\
 &+3\Gamma^{Q}_{T}\Gamma^{R}_{U}\,\tilde L'^{PQR}_{SUT}],
\label{eq:h1sf}\\
\frac{\partial\langle\hat{\mathcal H}_{\rm 3l}\rangle_{\rm 2C}}
     {\partial\Gamma^{QR}_{TU}}
 &=3\sum_{PS}\Gamma^{P}_{S}\,\tilde L'^{PQR}_{STU}.
\label{eq:h2sf}
\end{align}
The effective operator retains the form of Eq.~\eqref{eq:meanfield},
\begin{align}
\hat{\mathcal H}_{\rm 3l}\overset{\text{2C}}{\simeq}
 &\hat{\mathcal H}^{\rm (eff)}_{1}+\hat{\mathcal H}^{\rm (eff)}_{2}
  -\langle\hat{\mathcal H}_{\rm 3l}\rangle_{\rm 2C}
  +2\langle\hat{\mathcal H}_{\rm 3l}\rangle_{\rm 1C},\\
\langle\hat{\mathcal H}_{\rm 3l}\rangle_{\rm 1C}
 =&\sum_{PQRSTU}\Gamma^{P}_{S}\Gamma^{Q}_{T}\Gamma^{R}_{U}
   (\tilde L'^{PQR}_{STU}-\tfrac12\tilde L'^{PQR}_{SUT}).
\label{eq:meanfieldsf}
\end{align}

The corresponding traces of the spin-free cumulants are
\begin{align}
\mathrm{Tr}\{\Lambda^{(1)}\}&=N,
\label{eq:trlamsf1}\\
\mathrm{Tr}\{\Lambda^{(2)}\}&=\tfrac12\sum_{PQ}\Gamma^{P}_{Q}\Gamma^{Q}_{P}-N,
\label{eq:trlamsf2}\\
\mathrm{Tr}\{\Lambda^{(3)}\}&=\sum_{PQR}\Gamma^{P}_{Q}\Gamma^{Q}_{R}\Gamma^{R}_{P}
 -3\sum_{PQ}\Gamma^{P}_{Q}\Gamma^{Q}_{P}+2N .
\label{eq:trlamsf3}
\end{align}

We also derive the spin-free variant of the GBC correction for restricted open-shell Hartree--Fock (ROHF) and multiconfigurational SCF (MCSCF) references.
The object that replaces Eq.~\eqref{eq:gbc} is the generalized Fock matrix between the orbital space and its complement,
\begin{align}
F^{I}_{\alpha}&=F^{I}_{\alpha}[\hat h]+F^{I}_{\alpha}[\hat v],
\label{eq:genfock}\\
F^{I}_{\alpha}[\hat h]&=\sum_{J}\Gamma^{I}_{J}(0)\,\langle J|\hat h|\alpha\rangle,
\label{eq:genfockh}\\
F^{I}_{\alpha}[\hat v]&=\sum_{JKL}\Gamma^{IJ}_{KL}(0)\,G^{\alpha J}_{KL},
\label{eq:genfockv}
\end{align}
built from the RDMs of the reference $\Gamma(0)$ and vanishing in a complete basis.
Contracting each part with the geminal integral over $\alpha$ and $I$ and taking the expectation value over the remaining two indices, we obtain the two contributions to the residual,
\begin{align}
\langle\hat{\mathcal H}_{\rm 2h}\rangle_{0}
 &=\sum_{I\alpha}\sum_{JK}F^{I}_{\alpha}[\hat h]\,
   (R^{\alpha J}_{IK}-\tfrac12R^{\alpha J}_{KI})\,\Gamma^{J}_{K}(0),
\label{eq:e2hsf}\\
\langle\hat{\mathcal H}_{\rm 2v}\rangle_{0}
 &=\sum_{I\alpha}\sum_{JK}F^{I}_{\alpha}[\hat v]\,
   (R^{\alpha J}_{IK}-\tfrac12R^{\alpha J}_{KI})\,\Gamma^{J}_{K}(0),
\label{eq:e2vsf}
\end{align}
so that the correction retains the form of Eq.~\eqref{eq:dgbc},
\begin{align}
\Delta_{\rm GBC}=-\langle\hat{\mathcal H}_{\rm 2h}\rangle_{0}
                 -\langle\hat{\mathcal H}_{\rm 2v}\rangle_{0}.
\label{eq:hgbc}
\end{align}
The one-electron part is bilinear in $\Gamma^{(1)}(0)$, and the two-electron part is linear in $\Gamma^{(2)}(0)$.
At a restricted Hartree--Fock (RHF) reference, this formulation reduces to the spin-dependent correction of Sec.~\ref{sec:gbc}.

\section{Results and discussion}

In this section, we examine the performance of the pTC Hamiltonian numerically.
All calculations are performed using the \texttt{GELLAN} quantum chemistry program package.\cite{gellan}
To reduce the storage and computational cost, the cumulant implementation of pTC based on 3-electron integrals was converted into a grid-direct algorithm with the AI coding agent Claude Code (Anthropic, versions 2.1.197--2.1.266) running the Opus 5 model.
The same tool was used to analyze the output data, to speed up both tasks.
The geminal exponent is $\gamma=1.0\,a_{0}^{-1}$ throughout the paper, and the integrals are calculated by the numerical quadrature.\cite{tenno2004jcp}
Unless otherwise indicated, all results reported below are obtained within the spin-free framework of Sec.~\ref{sec:spinfree}.
The reference orbitals are those of the complete active space SCF (CASSCF),\cite{roos1980} and the correlation-consistent basis sets are those of Refs.~\onlinecite{dunning1989,kendall1992,woon1995} with the explicitly correlated set of Ref.~\onlinecite{peterson2008}.

\subsection{Main features of the effective mean field}
\label{sec:n2}

We first examine the accuracy of the 2C approximation using an exactly solvable model within the active space, the full-valence complete active space CI (CAS-CI) (10,8) for N$_2$ in the cc-pVTZ basis.
We employ the self-consistent mean-field (SCMF) in which the effective interaction is rebuilt from the relaxed density the FCI solver returns until convergence, and the orbitals are those of CASSCF in the same active space.
Our FCI solver treats a non-Hermitian Hamiltonian containing a 3-body operator explicitly, and the eigenvalue problem is solved by the generalization of the Davidson method to nonsymmetric matrices.\cite{hirao1982}

\begin{figure}[tb]
\includegraphics[width=\columnwidth]{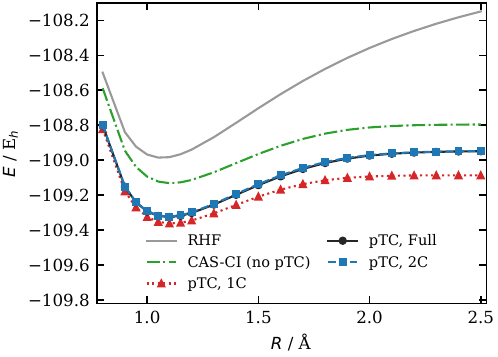}
\caption{Potential energy curves of N$_2$ in the cc-pVTZ basis, from CASSCF(10,8) orbitals and a CAS-CI in the $8$ active orbitals with the two $1s$ orbitals frozen.
\emph{Full} treats the linear 3-body operator of Eq.~\eqref{eq:h3l} explicitly; 1C and 2C replace it by the effective interactions of Eqs.~\eqref{eq:eff1sf} and~\eqref{eq:eff2sf}.
RHF and standard CAS-CI are given for reference.}
\label{fig:curves}
\end{figure}

Figure~\ref{fig:curves} shows the potential energy curves.
Against the experimental equilibrium dissociation energy $228.4$~kcal\,mol$^{-1}$,\cite{huber1979} the standard CAS-CI and that with pTC (CAS-pTC) with the full 3-body give $210.9$~kcal\,mol$^{-1}$ and $236.7$~kcal\,mol$^{-1}$, respectively.
pTC reduces the error in the dissociation energy by more than half in the same full-valence space.
The 2C curve is indistinguishable from the full one on the scale of the figure and returns $235.4$~kcal\,mol$^{-1}$ of dissociation energy, while the 1C approximation, evaluated at the same density, gives only $172.9$~kcal\,mol$^{-1}$.
All correlated curves give a minimum at $1.10$~{\AA}, against the experimental value of $1.0977$~{\AA}.

\begin{table}[b]
\caption{Basis-set convergence of the residual for N$_2$ at $R=1.10$~{\AA}, in mE$_{h}$.
The CASSCF(10,8) reference and the pTC space are held fixed --- only the AO basis changes, the number of frozen virtuals growing with it so that the pTC space remains $8$ orbitals.}
\label{tab:reslimit}
\begin{ruledtabular}
\begin{tabular}{lrrrr}
basis & $N_{\rm bf}$ & $\langle\hat{\mathcal H}_{\rm 2h}\rangle_{0}$
 & $\langle\hat{\mathcal H}_{\rm 2v}\rangle_{0}$ & $\Delta_{\rm GBC}$ \\
\hline
cc-pVDZ &  28 & 1172.92 & $-1159.54$ & $-13.38$ \\
cc-pVTZ &  60 & 1179.28 & $-1176.39$ &  $-2.89$ \\
cc-pVQZ & 110 & 1181.04 & $-1180.75$ &  $-0.29$ \\
cc-pV5Z & 182 & 1181.92 & $-1181.85$ &  $-0.07$ \\
\end{tabular}
\end{ruledtabular}
\end{table}

\begin{figure}[tb]
\includegraphics[width=\columnwidth]{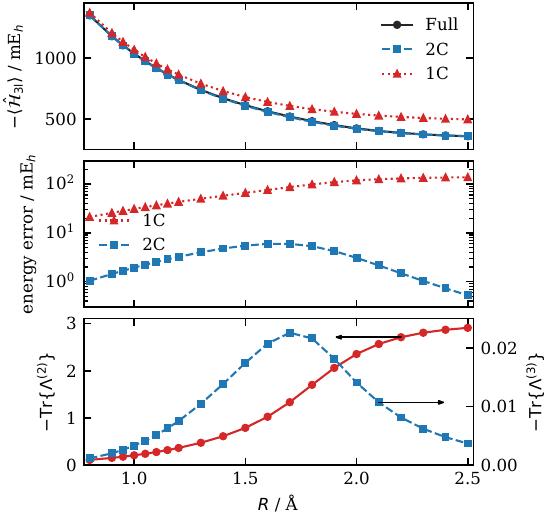}
\caption{The linear 3-body operator, the error of its cumulant approximations, and the traces of the discarded cumulants, along the dissociation of N$_2$ for the calculations of Fig.~\ref{fig:curves}.
Upper panel: $-\langle\hat{\mathcal H}_{\rm 3l}\rangle$ in full and in the 1C and 2C approximations, the latter two at the converged self-consistent density.
Middle panel: the 1C and 2C energy error relative to the full 3-body treatment, logarithmic scale.
Lower panel: $-\mathrm{Tr}\{\Lambda^{(2)}\}$ (left axis) and $-\mathrm{Tr}\{\Lambda^{(3)}\}$ (right axis) at the converged 2C density.
The upper and lower quantities are negative throughout and are plotted with the sign reversed.}
\label{fig:trace}
\end{figure}

Figure~\ref{fig:trace} attributes the error of each truncation, 2C or 1C, to the discarded cumulant.
The full 3-body energy $\langle\hat{\mathcal H}_{\rm 3l}\rangle$ ranges from $-1348$ to $-356$~mE$_{h}$, shrinking monotonically as the bond distance increases.
At the equilibrium bond length, this amounts to $\langle\hat{\mathcal H}_{\rm 3l}\rangle=-916.32$~mE$_{h}$.
The main contribution is from the two-electron part of the generalized Fock operator of Eq.~\eqref{eq:genfock}, $\langle\hat{\mathcal H}_{\rm 2v}\rangle_{0}=-1176.4$~mE$_{h}$, which is canceled by the one-electron part, $\langle\hat{\mathcal H}_{\rm 2h}\rangle_{0}=1179.3$~mE$_{h}$, as the GBC requires.
The remainder, $\langle\hat{\mathcal H}_{\rm 3l}\rangle-\langle\hat{\mathcal H}_{\rm 2v}\rangle_{0}=260.1$~mE$_{h}$, contributes positively to the total energy.
Table~\ref{tab:reslimit} traces $\Delta_{\rm GBC}$ as the basis is enlarged with the CASSCF reference and the pTC space held fixed.
$\Delta_{\rm GBC}$ falls monotonically by a factor of approximately $190$ from cc-pVDZ to cc-pV5Z, as the vanishing of Eq.~\eqref{eq:genfock} in a complete basis requires, although $\langle\hat{\mathcal H}_{\rm 2h}\rangle_{0}$ and $\langle\hat{\mathcal H}_{\rm 2v}\rangle_{0}$ themselves change by only a few mE$_{h}$.

The 2C error ranges from $1.0$ to $6.0$~mE$_{h}$, or $0.1$ to $1.2\%$ of the entire 3-body term, and peaks at $R=1.70$~{\AA} where the bond is half broken.
The 1C error is much larger, ranging from $21$ to $139$~mE$_{h}$, and grows monotonically with the bond length.
The traces of the discarded cumulants explain both behaviors.
$\mathrm{Tr}\{\Lambda^{(3)}\}$, absent in 2C, tracks the 2C error, peaking at the same $R=1.70$~{\AA}, and $\mathrm{Tr}\{\Lambda^{(2)}\}$, absent in 1C, grows monotonically across the range and follows the 1C error.
Physically, N$_2$ approaches a pair of open-shell atoms in the dissociation limit, where the pair correlation is strong while the connected three-particle correlation remains weak.
\enlargethispage{\baselineskip}

\begin{table*}[!t]
\caption{Error statistics against the HEAT reference, for the total energies (30 species, the H atom excluded), the atomization energies (26 molecules) and the reaction energies (16 reactions).
MAE is the mean absolute error and max is the largest absolute error.}
\label{tab:heat}
\begin{ruledtabular}
\begin{tabular}{lrrrrrrrrrrrr}
 & \multicolumn{6}{c}{aug-cc-pCVDZ} & \multicolumn{6}{c}{aug-cc-pCVTZ} \\
\cline{2-7}\cline{8-13}
 & \multicolumn{2}{c}{$\Delta E$ / mE$_{h}$} & \multicolumn{2}{c}{$D_{e}$ / kJ\,mol$^{-1}$} & \multicolumn{2}{c}{$\Delta H$ / kJ\,mol$^{-1}$} & \multicolumn{2}{c}{$\Delta E$ / mE$_{h}$} & \multicolumn{2}{c}{$D_{e}$ / kJ\,mol$^{-1}$} & \multicolumn{2}{c}{$\Delta H$ / kJ\,mol$^{-1}$} \\
\cline{2-3}\cline{4-5}\cline{6-7}\cline{8-9}\cline{10-11}\cline{12-13}
method & MAE & max & MAE & max & MAE & max & MAE & max & MAE & max & MAE & max \\
\hline
HF & 475.51 & 939.72 & 342.85 & 586.03 & 91.35 & 179.87 & 452.39 & 893.01 & 327.49 & 554.79 & 88.38 & 174.40 \\
MP2 & 171.42 & 322.58 & 44.81 & 90.93 & 21.82 & 42.85 & 71.34 & 121.44 & 18.02 & $-71.29$ & 12.23 & $-28.53$ \\
CCSD(T) & 148.47 & 300.82 & 71.37 & 134.53 & 18.35 & 33.49 & 47.75 & 97.19 & 23.33 & 47.44 & 5.91 & 10.42 \\
CCSD(T)-F12a & 38.32 & 81.65 & 13.30 & 26.08 & 3.65 & 7.24 & 7.71 & 16.23 & 2.86 & 5.82 & 0.79 & $-2.53$ \\
CCSD(T)-F12b & 44.26 & 90.28 & 17.30 & 28.19 & 5.30 & 10.40 & 13.08 & 25.61 & 5.41 & 9.07 & 1.63 & $-3.45$ \\
CCSD(T)-pTC & 53.02 & 111.66 & 72.67 & 146.20 & 15.54 & 36.19 & 13.95 & 37.94 & 13.02 & 50.08 & 5.74 & 18.99 \\
\quad $+\,\Delta_{\rm HF}$ & 22.57 & 50.41 & 55.98 & 111.24 & 13.31 & 31.32 & 6.62 & 23.40 & 11.79 & 46.36 & 5.47 & 18.31 \\
\quad $+\,\Delta_{\rm HF}+\Delta_{\rm GBC}$ & 6.98 & 14.23 & 14.10 & 25.90 & 5.63 & $-13.64$ & 3.60 & 13.32 & 5.93 & 24.34 & 3.18 & $-8.33$ \\
\end{tabular}
\end{ruledtabular}
\end{table*}

\begin{figure}[t]
\includegraphics[width=\columnwidth]{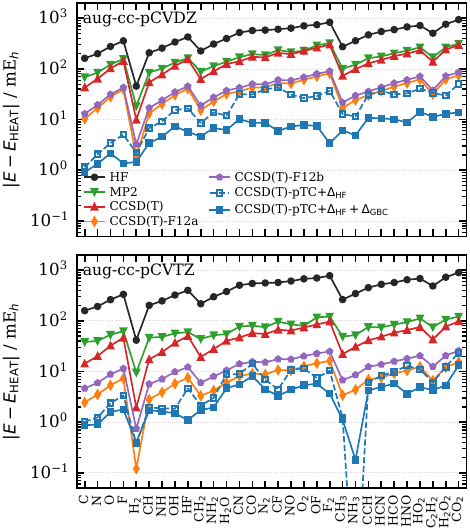}
\caption{Absolute total-energy error against the nonrelativistic limit of the HEAT protocol,\cite{tajti2004heat} species by species.
Upper panel: aug-cc-pCVDZ. Lower panel: aug-cc-pCVTZ.
All methods are all-electron in the same basis and geometry.
Species are ordered by size and the H atom is omitted, having no pTC step.
Open squares are pTC with $\Delta_{\rm HF}$ alone, filled squares add $\Delta_{\rm GBC}$.}
\label{fig:heat}
\end{figure}

\subsection{CCSD(T)-pTC}
\label{sec:heat}

We now turn to the assessment of CCSD(T)\cite{raghavachari1989} combined with pTC (CCSD(T)-pTC) on the $31$ atoms and molecules of the HEAT test set.\cite{tajti2004heat,bomble2006heat2,harding2008heat3}
The species other than the H atom are treated with RHF references for the $12$ closed shells and ROHF references for the $18$ open shells.
The use of SCMF is not practical for this model since RDMs span the full orbital space.
We therefore use the spin-free 2C mean field of Sec.~\ref{sec:spinfree} built at the reference density.
In this case, the 1C and 2C expectation values coincide at an RHF reference, and differ for ROHF only by the spin-free reduction of Eq.~\eqref{eq:g1half}.
The asymmetry of the integrals is retained by building the left and the right connected triples separately, while the denominators are diagonal, the $f_{ia}$ term is omitted, and $\Lambda$ is replaced by $T$.
The resulting error in the total energy, measured against the CCSD relaxed density, is $0.6$~mE$_{h}$ on average and at most $1.3$~mE$_{h}$ (CO$_{2}$) in aug-cc-pCVDZ, and falls to $0.08$~mE$_{h}$ in aug-cc-pCVTZ.
The comparison is against the nonrelativistic limit of the HEAT set, {\it i.e.}\ the total energy without its relativistic, zero-point, diagonal Born--Oppenheimer and spin--orbit terms.
The CCSD(T)-F12 calculations are carried out with \texttt{MOLPRO},\cite{molpro} all electrons correlated, using the F12/3C(FIX) ansatz with a Slater geminal of exponent 1.0, the cc-pCVQZ-F12/JKFIT set as the complementary auxiliary basis, and cc-pCVQZ-F12/MP2FIT and cc-pCVQZ-F12/JKFIT for the Coulomb and exchange density fitting.

Figure~\ref{fig:heat} shows the total-energy error of each method, and Table~\ref{tab:heat} collects the statistics for the total energies, the $26$ atomization energies and the $16$ reaction energies.
We take two corrections into account alongside the pTC energy.
$\Delta_{\rm HF}=E_{\rm SCF}({\rm CVQZF12})-E_{\rm SCF}$ repairs the basis-set incompleteness of the reference function, in place of the CABS singles correction.
$\Delta_{\rm GBC}$ is the residual of Eq.~\eqref{eq:hgbc}.
In aug-cc-pCVDZ, the MAE of the total energy falls from $53.0$ to $22.6$~mE$_{h}$ with $\Delta_{\rm HF}$ and to $7.0$~mE$_{h}$ with both $\Delta_{\rm HF}$ and $\Delta_{\rm GBC}$, against $38.3$~mE$_{h}$ for CCSD(T)-F12a and $148.5$~mE$_{h}$ for CCSD(T).
The pTC Hamiltonian is non-Hermitian and a smaller total-energy error does not imply a better description.
The relative energies are the more meaningful test.

In the atomization and reaction energies the F12 methods are more accurate by a small margin.
At aug-cc-pCVDZ, the corrected pTC atomization error, $14.1$~kJ\,mol$^{-1}$, falls between those of F12a and F12b, and at aug-cc-pCVTZ, F12 slightly outperforms pTC.
This reversal from the total-energy result originates in the character of the two errors.
The F12 error is nearly constant per atom and largely cancels in a difference of energies, its atomization error being about a third of its total-energy error.
The pTC error does not cancel, the atomization error exceeding the total-energy error in both bases.
$\Delta_{\rm HF}$ removes more than half of the pTC total-energy error but less than a quarter of the atomization error, whereas $\Delta_{\rm GBC}$ removes $75\%$ and $50\%$ of the atomization error at double- and triple-zeta.
Without $\Delta_{\rm HF}$ and $\Delta_{\rm GBC}$, the pTC energy is no more accurate than CCSD(T) for the atomization energies
($72.7$ against $71.4$~kJ\,mol$^{-1}$ at aug-cc-pCVDZ), and $\Delta_{\rm GBC}$ in particular brings pTC to F12 accuracy.
The 3-body term is modest, $\langle\hat{\mathcal H}_{\rm 3l}\rangle_{\rm 2C}$ averaging $158$~mE$_{h}$ in absolute value over the aug-cc-pCVDZ set and reaching $391$~mE$_{h}$ for F$_{2}$.
At worst with $1\%$ truncation error, this is a few~mE$_{h}$ at most.
Although the result of the spin-free formulation is tabulated, the spin-dependent result reproduces every entry to within $1\%$, where the difference is only for the open-shell species.

\begin{figure}[t]
\includegraphics[width=\columnwidth]{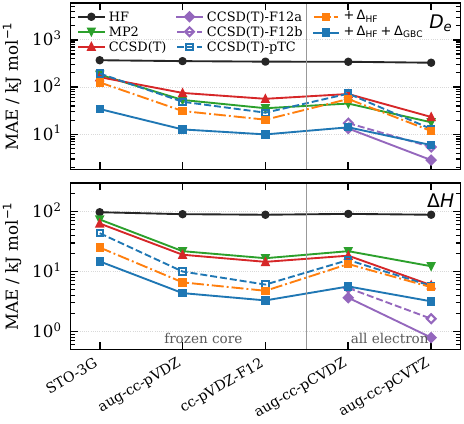}
\caption{Mean absolute error of the atomization energies (upper) and reaction energies (lower) against the HEAT reference, as the one-particle basis is enlarged.
The three bases left of the rule are frozen core and the two on its right all electron, so the abscissa is not a single basis-set series.
HF, MP2 and CCSD(T) are the companion calculations without pTC for the frozen-core sets, and the values of Fig.~\ref{fig:heat} for the all-electron ones.
$\Delta_{\rm HF}$ and $\Delta_{\rm GBC}$ are as in Fig.~\ref{fig:heat}.
The F12 values are those of Table~\ref{tab:heat} and are shown for the all-electron sets only.}
\label{fig:basis}
\end{figure}

Figure~\ref{fig:basis} depicts the two relative quantities across five one-particle basis sets, three of them frozen core and two all electron.
With both corrections the pTC error is the lowest at every frozen-core basis.
At cc-pVDZ-F12, the error is reduced to $10.0$~kJ\,mol$^{-1}$ for the atomization energies and $3.3$~kJ\,mol$^{-1}$ for the reaction energies, which the standard CCSD(T) does not reach even at aug-cc-pCVTZ ($23.3$ and $5.9$~kJ\,mol$^{-1}$).
With the 2C mean field in place of the 3-body operator, pTC and the two corrections therefore let a double-zeta basis designed for explicit correlation outperform the standard CCSD(T) in a triple-zeta basis.

\begin{figure}[tb]
\includegraphics[width=\columnwidth]{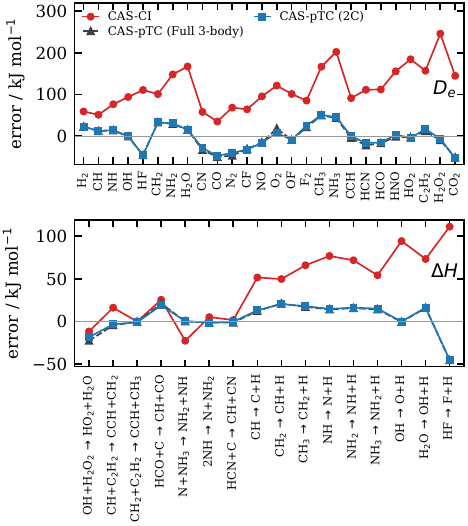}
\caption{Error of the atomization energies (upper) and reaction energies (lower) in the full-valence CAS space alone, against the HEAT reference.
CASSCF/aug-cc-pVTZ orbitals, $1s$ frozen.
CAS-CI is the standard CI in that space.
The two pTC curves are the same CI on the pTC Hamiltonian, with the 3-body operator either treated explicitly or replaced by the 2C mean field of Sec.~\ref{sec:spinfree}.}
\label{fig:cas}
\end{figure}

\begin{table}[b]
\caption{Error against the HEAT reference of the atomization energies (26 molecules) and reaction energies (16 reactions) obtained in the full-valence CAS space alone, with CASSCF/aug-cc-pVTZ orbitals and the $1s$ pairs frozen, in kJ\,mol$^{-1}$.
\emph{Full} denotes the 3-body operator treated explicitly and \emph{sd} the spin-dependent 2C mean field of Sec.~\ref{sec:cumulant}.}
\label{tab:cas}
\begin{ruledtabular}
\begin{tabular}{lrrrr}
 & \multicolumn{2}{c}{$D_{e}$} & \multicolumn{2}{c}{$\Delta H$} \\
\cline{2-3}\cline{4-5}
 & mean & MAE & mean & MAE \\
\hline
CAS-CI & 116.0 & 116.0 & 41.4 & 45.7 \\
CAS-pTC, Full & -2.4 & 23.5 & 3.4 & 12.9 \\
CAS-pTC, Full/Real & -3.5 & 24.1 & 3.2 & 13.0 \\
CAS-pTC, 2C/Real & -2.1 & 23.2 & 3.7 & 12.7 \\
CAS-pTC, 2C(sd)/Real & -1.2 & 23.3 & 3.5 & 13.1 \\
\end{tabular}
\end{ruledtabular}
\end{table}

\subsection{The downfolded Hamiltonian on a qubit register}

We now consider qubit Hamiltonians obtained by the pTC downfolding.
For this purpose, we use the CAS-pTC model which typically consists of a CAS-CI in the full-valence active space with the CASSCF orbitals of the same space.
The correlation outside the active space is treated only through the correlation factor.
A near-term or early fault-tolerant register can hold this structure, since only the active space is mapped onto the qubits.
The $31$ species of the HEAT set are treated again with the aug-cc-pVTZ basis and the $1s$ pairs frozen.
The mean absolute value of $\Delta_{\rm GBC}$ over these $31$ species is $1.8$~mE$_{h}$.
The results below are reported without it, in anticipation of wider applications in which the register holds the operator alone and no correction evaluated from the reference function is available.

The register receives the operator through the Jordan--Wigner mapping,\cite{jordan1928}
\begin{align}
\ad{p}&=\prod_{q<p}\hat Z_{q}\,\frac{\hat X_{p}-i\hat Y_{p}}{2},
\label{eq:jwc}\\
\an{p}&=\prod_{q<p}\hat Z_{q}\,\frac{\hat X_{p}+i\hat Y_{p}}{2},
\label{eq:jwa}
\end{align}
under which any operator becomes a sum over Pauli strings $\hat P_{k}$ that separates into Hermitian and anti-Hermitian parts,
\begin{align}
\hat{\mathcal H}=\hat{\mathcal H}_{\rm H}+\hat{\mathcal H}_{\rm A},
\label{eq:pauliexp}
\end{align}
with
\begin{align}
\hat{\mathcal H}_{\rm H}&=\sum_{k}c_{k}\,\hat P_{k},
\label{eq:pauliherm}\\
\hat{\mathcal H}_{\rm A}&=i\sum_{k}d_{k}\,\hat P_{k}.
\label{eq:pauliantiherm}
\end{align}
Each $\hat P_{k}$ is Hermitian, and with real integrals its coefficient is real or imaginary according to the parity of the number of $\hat Y$ factors, so the two sums run over disjoint sets of strings.
The pTC Hamiltonian contains anti-Hermitian components and its expansion is complex.
Retaining both parts is referred to below as the complex expansion, and retaining $\hat{\mathcal H}_{\rm H}$ alone as the real expansion.

Figure~\ref{fig:cas} shows the error of each species, and Table~\ref{tab:cas} lists the errors over the set, for the CAS-CI and for CAS-pTC with the 3-body operator either treated explicitly or replaced by the 2C SCMF.
The standard CAS-CI underestimates the binding throughout, by $116$~kJ\,mol$^{-1}$ on average in the atomization energies and $41$~kJ\,mol$^{-1}$ in the reaction energies.
This deficit is due to the dynamic correlation that the active space cannot accommodate.
Solving the same CI on the pTC Hamiltonian recovers most of it, the mean error in the atomization energies falling to $-2$~kJ\,mol$^{-1}$ and the mean absolute errors to $23$ and $13$~kJ\,mol$^{-1}$, without any enlargement of the active space.
Table~\ref{tab:cas} then separates the two approximations a register requires, the real expansion and the 2C mean field.

\begin{figure}[t]
\includegraphics[width=\columnwidth]{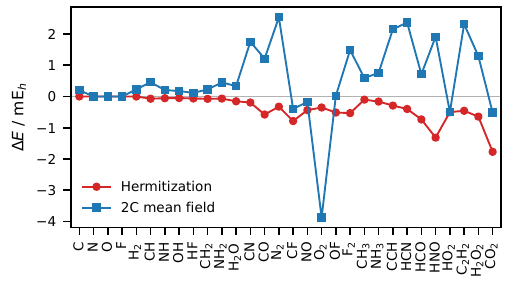}
\caption{The errors of the two approximations in the total energy.
The Hermitization error is $E$(3-body, real)$-E$(3-body, complex) and the 2C mean field one is $E$(2C, real)$-E$(3-body, real), in the CAS space of Table~\ref{tab:cas}.
The species are ordered as in Fig.~\ref{fig:heat}.}
\label{fig:eff}
\end{figure}

\begin{table}[t]
\caption{Number of distinct Pauli strings, identity included, in the Jordan--Wigner image of the standard CAS-CI Hamiltonian and of CAS-pTC, over $M$ spatial orbitals on $N=2M$ qubits.
The point group is the abelian subgroup used in the calculation.
The complex expansion retains both parts of Eq.~\eqref{eq:pauliexp} and the real expansion only $\hat{\mathcal H}_{\rm H}$.
Only the strings actually generated, with a coefficient exceeding $10^{-14}$, are counted.}
\label{tab:pauli}
\begin{ruledtabular}
\begin{tabular}{cccrrrr}
 & & & CAS-CI & \multicolumn{3}{c}{CAS-pTC} \\
\cline{5-7}
Molecule & qubit & group & & 3-body & 3-body & 2C \\
 & & & & complex & real & real \\
\hline
H$_{2}$O   & 12 & $C_{2v}$ &   551 &  18\,459 &   9\,379 &     823 \\
CH$_{3}$   & 14 & $C_{2v}$ & 1\,086 &  52\,526 &  26\,498 &  1\,646 \\
NH$_{3}$   & 14 & $C_{s}$  & 1\,742 &  98\,954 &  49\,714 &  2\,686 \\
N$_2$    & 16 & $D_{2h}$ &   825 &  58\,109 &  29\,449 &  1\,313 \\
CO         & 16 & $C_{2v}$ & 1\,617 & 116\,933 &  58\,841 &  2\,505 \\
HCN        & 18 & $C_{2v}$ & 2\,740 & 252\,668 & 126\,892 &  4\,284 \\
H$_{2}$O$_{2}$ & 20 & $C_{2}$ & 7\,151 & 981\,891 & 491\,631 & 11\,691 \\
CO$_{2}$   & 24 & $D_{2h}$ & 4\,173 & 782\,117 & 392\,309 &  6\,813 \\
\end{tabular}
\end{ruledtabular}
\end{table}

A quantum circuit evolves a state in real time under a unitary transform.
The generator of the evolution must therefore be Hermitian, and the register admits only the real expansion.
A non-Hermitian generator can still be simulated through a linear combination of unitaries or a block encoding,\cite{childs2012,low2017} at the price of ancillae and a success probability.
The usability of the real expansion is therefore crucial for practical quantum computing.
Figure~\ref{fig:eff} separates the two approximations on the total energy.
Discarding the imaginary coefficients lowers every total energy, by $0.41$~mE$_{h}$ on average over the $31$ species and at most $1.77$~mE$_{h}$, for CO$_{2}$, while leaving an atom unaffected to within $10^{-4}$~mE$_{h}$.
The lowering grows with the number of electrons, whereas the shift from the 2C mean field changes sign, between $-3.9$ and $2.5$~mE$_{h}$.
In the relative energies, the Hermitization by the real expansion slightly increases the atomization error from $23.5$ to $24.1$~kJ\,mol$^{-1}$ and the reaction error from $12.9$ to $13.0$.
Replacing the 3-body operator by the 2C mean field hardly affects the MAEs, $23.2$ and $12.7$~kJ\,mol$^{-1}$ against the $24.1$ and $13.0$ of the real 3-body operator.
The 2C approximation is therefore sufficiently accurate.
The spin-dependent 2C of Sec.~\ref{sec:cumulant} is indistinguishable from the spin-free result over the set, $23.3$ and $13.1$~kJ\,mol$^{-1}$ in the last row, but dependent on species.
The 2C error is largest for O$_{2}$, $10.2$~kJ\,mol$^{-1}$ in the atomization energy, but in fact the spin-dependent form gives only $-2.9$~kJ\,mol$^{-1}$.
The reason is that Eq.~\eqref{eq:g1half} holds for the $M_{S}$-averaged ensemble and not for the calculated $M_{S}=1$ state.
The excess error therefore belongs to the spin-free reduction and not to the truncation of the cumulant.

The greatest advantage of 2C on the register is the reduction in the number of terms to be measured.
Counting each distinct Pauli string once, the number of strings scales as $M^{4}$ for a spin-conserving operator of rank at most two and as $M^{6}$ once the 3-body term of Eq.~\eqref{eq:h3l} is present, whatever the ordering of the register.
Table~\ref{tab:pauli} lists the counts for eight species, from 12 to 24 qubits.
Both Hamiltonians are constructed in the CASSCF natural orbitals canonicalized with the generalized Fock operator without pTC.
In this basis the one-body part of the pTC Hamiltonian is less sparse than that of the bare Hamiltonian, and the CAS-CI counts are smaller than those of CAS-pTC 2C.
Against the standard CAS-CI Hamiltonian in the same space, CAS-pTC with the 2C mean field costs a factor of $1.5$ to $1.6$ in strings, while the explicit 3-body costs a factor that grows from $17$ at $N=12$ to $94$ at $N=24$.
The downfolded operator is therefore measured at close to the price of the Hamiltonian it replaces, and the margin over the untruncated form widens with the system.

\section{Conclusions}
In this paper, the effective interactions have been derived by partial differentiation of the 2C expectation value of the cumulant approximation, which replaces the 3-body term of the pTC Hamiltonian by up to 2-body operators.
The GBC correction has been examined, and both it and the effective interactions have been extended to spin-free formulations.

The truncation of 3-electron integrals with the 2C approximation has been assessed along the dissociation of N$_2$, where the trace of the discarded cumulant accounts for the error.
On the HEAT set, CCSD(T)-pTC with the $\Delta_{\rm HF}$ and $\Delta_{\rm GBC}$ corrections is more accurate in every double-zeta basis examined, frozen core and all electron, than the standard CCSD(T) at triple zeta.
A real-time evolution requires a Hermitian generator, and the register restricted to the real expansion slightly degrades the accuracy.
The 2C mean field affects the relative energies very little and reduces the number of Pauli strings from the sixth power of the orbital count to the fourth.

Several questions still remain.
The terms $\hat{\mathcal H}_{\rm 3q}$ and $\hat{\mathcal H}_{\rm 4}$ of Eq.~\eqref{eq:hptc}, which have not been examined, are unlikely to be negligible in a small orbital space.
A more robust formulation of pTC is also conceivable by analogy with the CC Lagrangian as
$\hat{\mathcal H}'_{\rm pTC}
   =\mathcal B\{\hat{\mathcal G}^\dagger e^{-\hat{\mathcal G}}\,\hat H\,e^{\hat{\mathcal G}}\}$,
which is still terminating.
There is also room for an effective treatment of the long-range correlation that CAS-pTC leaves outside the active space, either by introducing more flexibility in the correlation factor or by using auxiliary functions.
For excited states, the 2C approximation can be made state-specific, although our preliminary investigation indicates that the use of the ground-state density closely reproduces the low-lying excitation spectrum of the full 3-body treatment.

The CAS-pTC model developed here has been used in several quantum algorithms, as will be reported elsewhere.

\begin{acknowledgments}
The author is grateful to Daniel Kats for his valuable comments on an earlier version of this manuscript.
This work was partially supported by MEXT as ''Program for Promoting Researches on the Supercomputer Fugaku'' (Realization of innovative light energy conversion materials, Grant Number JPMXP1020210317), and JSPS KAKENHI Grant 22H00316. 
\end{acknowledgments}

\section*{AUTHOR DECLARATIONS}
\subsection*{Conflict of Interest}
The author has no conflicts to disclose.

\subsection*{Author Contributions}
Seiichiro Ten-no: Conceptualization (equal); Data curation (equal); Formal analysis (equal); Funding acquisition (equal); Investigation (equal); Methodology (equal); Project administration (equal); Resources (equal); Software (equal); Supervision (equal); Validation (equal); Visualization (equal); Writing -- original draft (equal); Writing -- review \& editing (equal).

\section*{DATA AVAILABILITY}
The data that support the findings of this study are available from the corresponding author upon reasonable request.

\bibliographystyle{aipnum4-1}
\bibliography{ptc_cumulant}

\end{document}